# Value of the Teaching Career and Factors in Its Path in Peru

**Michelle Lozada-Urbano [a*], Delsi Mariela Huaita Acha [b], Rosa Maria Benavente Ayquipa [c], Freddy Felipe Luza Castillo [d], Rosse Mary Falcón-Antenucci [e]**

[a] Vice-Rectorate for Research, Norbert Wiener University, Av. Arequipa 440, Lima 15073, Lima, Peru, michelle.lozada@uwiener.edu.pe
[b] Inca Garcilaso de la Vega University, Av. Arequipa 3690, San Isidro 15088, Lima, Perú, delsi.huaita@uigv.edu.pe
[c] San Marcos National University, Av. Universitaria s/n, cruce con Av. Venezuela, Lima15081, Lima, Peru, rbenaventea@unmsm.edu.pe
[d] Cesar Vallejo University; Av. Alfredo Mendiola 6232, Los Olivos 15306, Lima, Peru, fluza@ucvvirtual.edu.pe
[e] San Juan Bautista University, Av. San Luis 1923, San Borja 15037, Lima, Perú, rosse.falcon@upsjb.edu.pe

**Corresponding author:**
Michele Lozada-Urbano, Vice-Rectorate for Research, Norbert Wiener University, Av. Arequipa 440, Lima, Peru; 15073, e-mail: michelle.lozada@uwiener.edu.pe

**Declaration of interests**
The authors declare that they have no known competing financial interests or personal relationships that could have appeared to influence the work reported in this paper.

**Abstract**

The teaching career shares common global characteristics, such as internal promotion, performance evaluation, recruitment of top candidates, continuous training, specialization, and peer learning. This study aims to describe the factors associated with the value placed on the teaching career in Peru. A total of 28,217 public school teachers were analyzed using data from the 2020 National Teacher Survey. A variable measuring the "value of the teaching career" was constructed using eight indicators and categorized as low, medium, or high. Another variable, "vision of the future," was classified as pessimistic, conformist, or optimistic. This observational, cross-sectional, and analytical study included variables related to in-service training, working conditions, professional recognition, and sociodemographic characteristics. Among the teachers surveyed, 45.8% expressed an optimistic outlook on the future of the profession, 48% held a conformist view, and only 6.2% reported a pessimistic perspective. A generalized linear model revealed that the value placed on the teaching career was significantly associated with male gender ($p=0.002$), professional career ($p<0.001$), an optimistic outlook ($p=0.033$), and working at the primary level ($p<0.001$). It was concluded that, Peruvian teachers predominantly hold conformist or optimistic views of their profession. This highlights the need to reinforce merit-based advancement, competency-based training, intrinsic motivation, and ongoing professional development

**Highlights**

- Most Peruvian teachers tend to hold either conformist or optimistic perspectives about their profession
- Determinants of career appreciation
- Training and working conditions varied by career level
- Need for reinforced policies and support

# 1. Introduction

The year 2020 marked a significant shift in the education systems of both Peru and the world, as schools were temporarily closed to prevent the spread of COVID-19. This highlighted the lack of access to technology and the digital divides teachers faced, generating tension, worry, anguish, and anxiety when faced with a different work system (Minedu, 2021). According to the Endo test (2018), conducted before the closure of schools, 20% of teachers were from rural areas, 15% did not have a computer or laptop, 40% had no internet connection at home, and in rural areas, the figure rose to 60% (Minedu, 2018).

In 2020, with the temporary closure of schools, strategies were adopted to maintain educational services, such as online education, the distribution of technological devices, and packages of printed materials, so that those without access to technology could have the necessary tools to continue their education, generating dissatisfaction among many of them.

Every profession has its own sense of satisfaction or dissatisfaction in its development. Teacher devaluation is a long-standing issue. In 2016, the Ministry of Education described teachers as facing an identity crisis and struggling to gain social recognition and demonstrate that their profession is relevant and scientific. According to the National Education Project (PEN), in 2021, teachers were demotivated and lacked confidence in policies for change.

In countries like the United States, the devaluation of the profession is high, with 91% of upper-level students not considering pursuing a teaching career because they consider it unattractive due to its disrepute and image (Minedu, 2016). In countries with a higher value on teaching, such as Finland, South Korea, and Singapore, state policies focused on recruiting top students to study for the teaching profession, supporting their development, and implementing support mechanisms with quality and excellence in teaching have allowed for their social revaluation. In Singapore, the education system managed to have the best-performing schools, achieving high scores on the International Student Assessment (PISA). However, in Peru, scores on these international tests were already low before the lockdown.

During the first wave of the COVID-19 pandemic, teachers in Peru experienced mixed feelings about their satisfaction with their work as educators. This perception not only reflected how they adapted to new teaching requirements but also directly affected students' learning and development. Students, in turn, faced major challenges due to sudden changes in the educational system and daily routines, which required a complex readjustment (Ministry of Education, 2020).

The teaching career in Peru involves a promotion process with two main stages. The first stage prioritizes labor rights and ensures job stability, with the state acting in a protective capacity. The second stage focuses on meritocracy, motivating teachers to advance and seek better job opportunities, with the goal of improving learning outcomes.

To conceptualize the evaluation of the teaching career, it is essential to consider that assessment, performance recognition, and the quality of the educational service all contribute directly to student benefit. The primary goal of career evaluation is to measure and assess the work carried out, identifying both strengths and areas for improvement, while also rewarding those who demonstrate excellence. Some authors analyze the perception of the social valuation of the teaching career, describing associated aspects such as the level of satisfaction, professional practice, in-service training, management of illness, and its impact on physical and mental health (Gonzáles & Sánchez, 2021).

Some relevant aspects for assessing the teaching career are: a) Performance evaluation that guarantees pedagogical innovation and the constant assessment of knowledge. b) Recognition and rewards for teachers who demonstrate outstanding performance. c) Including professional development programs or in-service training. d) Clear standards and criteria in evaluation that guarantee fairness and reward merit. e) Constructive feedback that contributes to professional growth. f) Responsibility and accountability to ensure that teachers fulfill their duties and that measures are taken when performance is not optimal.

There are theories that support the assessment and professional development applied to teaching, such as the theory of professional capital, which seeks to ensure that the best professionals, with a high level of commitment, are those responsible for early childhood and basic education. This theoretical position maintains that the profession should be promoted, valued, and consolidated through training, incentives, good salaries, and in-service training that allow teachers to improve their skills (Quintero Montaño, 2020).

Human capital theory has been used in various professions, and applying it to the assessment of teaching careers proposes strengthening their skills by increasing their potential for the direct benefit of students and the community at large. Hence, the importance of the proposed analysis. The quality of education, improved job opportunities, ongoing training, and the retention of the best teachers are all important for the overall development of communities (Tantawy, 2020).

Spencer et al. (2018) analyzed the professional development needs of teachers in England, concluding that, despite the distancing of local authorities from schools, early career teachers were willing to participate in development activities provided by the government, although they felt mistreated and little support for the mission and beliefs of their profession. Kwee (2020) considered that the teacher shortage is due to demotivation due to personal and contextual factors or burnout due to gaps in research on the teaching profession.

In Latin American and Caribbean countries such as Colombia, Guatemala, Costa Rica, Mexico, Nicaragua, Paraguay, and the Dominican Republic, among others, teachers adhere to similar educational policies where meritocracy, in-service training, and attracting the best candidates for the teaching career are the foundation of their educational systems (Escribano, 2017).

Teacher satisfaction in Peru is closely linked to organizational climate, emotional well-being, and professional performance. Factors such as stress, lack of recognition, and difficulties accessing promotions contribute to teacher dissatisfaction. Conversely, a supportive work environment and effective administrative strategies can improve satisfaction, leading to improved performance (Parrales & Puero, 2021).

## 2. Materials and Methods

*2.1. Design and participants*

This research was an observational, cross-sectional, analytical study using secondary data. The study population was 28,217 teachers, corresponding to the total number of Regular Basic Education (EBR) teachers surveyed as part of the 2020 National Teacher Survey (Endo, 2020). The sampling method was census-based, as it collected data from the entire teacher population. The National Teacher Survey (Endo) was conducted by the Ministry of Education (MINEDU), which is conducted every two years. The data are publicly available at: (http://www.minedu.gob.pe/politicas/docencia/encuesta-nacional-a-docentes-endo.php).

The Endo 2020 survey collects information from Peruvian teachers regarding their sociodemographic and economic characteristics, their training and professional career, their perceptions of how they feel and the conditions under which they work, as well as the programs offered by the Ministry of Education (Minedu) to improve teaching and their outlook on the future. The Endo 2020 survey was administered in Peru's 26 regions, consisting of data from 28,217 teachers, 17,406 from urban areas, and 10,811 from rural areas. The information was collected through telephone calls. It provides inferences at the national, regional, and geographic area levels.

*2.2. Instrument and Variables*

*2.2.1. Design and Organization of Variables.*

Records with missing information exceeding 10% of the total were eliminated. Records with less than 10% missing information, or where the data were implausible, were considered missing data.

*2.2.2. Construction of the instrument to measure the valorization of the teaching career and its validation.*

The study technique was observation, and the instrument was developed based on variables existing in the Endo 2020 database. Initially, the three education researchers were responsible for developing an initial proposal for the instrument, after reviewing the variables available in the Endo 2020 database and identifying the characteristics that, according to theoretical constructs, would allow for exploring the valorization of the teaching career.

The research team analyzed the initial proposal for the instrument, both in terms of its ability to identify aspects of teacher valorization and in terms of how to score the responses and the categories those scores would produce. Finally, based on the research team's evaluation, the final version of the instrument was agreed upon. It contains eight aspects: in-service training, employment status, occupation outside the workplace, type of occupation outside the workplace, psychological support, COVID-19 experience of a family member, encouragement that teachers should receive, and participation in a project.

Each aspect was scored as follows:
- In-service training: virtual course (1 point), conferences/seminars and ICT training (2 points), teacher program (3), and no training (0 points).
- Employment status: contracted through another modality (1), contracted through a competitive examination (2), and appointed (3 points).
- Occupation outside the workplace: yes (2 points), no (3 points).
- Additional occupation: teacher at a private IE (1 point), teacher at a private tutoring academy (2 points), and other activities (agriculture, self-employed business, etc.) (2 points).
- Psychological support: if "2" points, not "1" point.
- COVID-19 experience of a family member: if "1" point, not "3" points.
- Incentives Teachers should receive: resolutions, bonuses, and scholarships: "1" point.
- Participation in a project: if "3" points, not "1" point.

The sum of the scores generated by each item was divided into: 1 to 14 points, 15 to 18 points, and 19 to 21 points. The score ranges generate three categories: low, medium, and high appreciation, respectively. However, for the analysis of the strength and direction of the association, the appreciation of the teaching career was dichotomized, establishing two categories based on the three previous categories: adequate (19 to 21 points) and inadequate (1 to 18 points).

Additionally, the Future Vision variable was constructed from the item "Job Future for 2025," which consists of three categories: pessimistic vision, conformist vision, and optimistic vision. These categories were constructed as follows: "pessimistic vision" from the items "leaving the teaching career and retiring early," "conformist vision" from the items "continuing as a teacher and retiring within the regular timeframe," and "optimistic vision" from the items "holding management positions, working at the Ministry of Education (Minedu), the Regional Directorate of Education (DRE), or the Local Educational Management Unit (UGEL).

*2.3. Data processing.*

A univariate analysis was conducted, processing the age variable with mean and standard deviation, while categorical variables were analyzed using frequencies and percentages. A bivariate analysis was performed on characteristics such as age, sex, career trajectory, future vision, and educational level, using the Chi-square statistic or Fisher's exact test when the Chi-square assumptions were not met.

Subsequently, a regression analysis was conducted using a generalized linear model (GLM) with a Poisson family and a log link function to obtain the direction and strength of association, expressed in prevalence ratios (PR), in both the crude (cPR) and adjusted (aPR) models. In this analysis, the appreciation of the teaching career was treated as a dichotomous variable, with categories of adequate or inadequate appreciation.
Statistical decisions were made considering a p-value <0.05 and a 95% confidence level. Statistical analysis was performed using STATA version 17.0.

*2.4. Ethical considerations*.

Secondary data were used; it was not necessary to submit the project for Ethics Committee approval.

## 3. Results

This result corresponds to the National Survey for Peruvian Teachers, conducted in 2020 (Endo, 2020) with a sample of 28,217 teachers. Table 1 describes the profile of the teachers, specifying their sociodemographic characteristics. The average age is 45.4 years; the highest percentage is categorized in the 40-59 age group; 67.4% are female; the largest group of respondents works at the primary level, followed by preschool, and finally, secondary school.

**Table 1.** Sociodemographic characteristics of teachers according to Endo 2020

| Characteristic | | N (%) |
|---|---|---|
| Age (years) | | 45.4 ± 0.069 |
| Categorized age | 20 to 39 | 5583 (29.5) |
| | 40 to 59 | 12024 (63.4) |
| | 60 to 70 | 1347 (7.1) |
| Gender | Female | 12786 (67.4) |
| | Male | 6173 (32.6) |
| Educational level | Preschool | 7091 (28.0) |
| | Primary | 11955 (42.4) |
| | Secondary | 8360 (29.6) |
| Career level | One | 3949 (37.81) |
| | Two | 3064 (29.34) |
| | Three | 2056 (19.69) |
| | Four | 873 (8.36) |
| | Five | 386 (3.70) |
| | Six | 107 (1.02) |
| | Seven | 8 (0.08) |

Table 2 shows that 10,443 teachers were appointed to the EBR (Educational Training) through a public competition, while 8,408 were hired through a competitive competition. A total of 86.93% of teachers reported having no additional occupation that generates income. Among those who reported having an additional occupation, those at level 1 were employed. 20% worked in a private institution and 23% gave private lessons. 46.5% of teachers without a teaching degree worked in agriculture, were laborers, or owned a business. 53.9% gave private lessons, 38.6% taught in high schools or universities, and 43.1% worked in private educational institutions.

Teachers without a scale or at level 1 consider that they should receive congratulatory resolutions, monetary bonuses, scholarships, and training provided by the Ministry of Education. 40.7% of teachers without a scale participated in pedagogical innovation projects, and 47.6% did not participate, with teachers at the highest scales being those who had the least participation.

Table 2. Teachers' job characteristics according to their level on the Teaching Scale

| Characteristic | Total N | Level on the teaching scale | | | | |
|---|---|---|---|---|---|---|
| | | No scale | I-II N (%) | III N (%) | IV N (%) | V-VII N (%) |
| **In-service training (N=10,434)** | | | | | | |
| Virtual Courses | 15996 | 7183 (44.9) | 5852 (36.6) | 1755 (11.0) | 751 (4.7) | 455 (2.8) |
| Conferences or seminars | 10615 | 4812 (45.3) | 3803 (35.8) | 1148 (10.8) | 517 (4.9) | 335 (3.2) |
| ICT Training | 12852 | 5937 (46.2) | 4602 (35.8) | 1366 (10.6) | 592 (4.6) | 355 (2.8) |
| "Private to Public IE absorption" Program | 1169 | 532 (45.5) | 461 (39.4) | 118 (10.1) | 36 (3.1) | 22 (1.9) |
| None | 1145 | 515 (45.0) | 464 (40.6) | 118 (10.3) | 42 (3.7) | 6 (0.5) |
| **Working Conditions and Outlook** | | | | | | |
| Additional occupation that generates economic income | | | | | | |
| Yes | 2473 | 1150 (46.5) | 910 (36.8) | 255 (10.3) | 104 (4.2) | 54 (2.2) |
| No | 16455 | 7343 (44.6) | 6097 (37.0) | 1800 (10.9) | 768 (4.7) | 447 (2.7) |
| Teacher's additional occupation (N=10,443) | | | | | | |
| In private IE | 130 | 56 (43.1) | 42 (32.3) | 19 (14.6) | 8 (6.2) | 5 (3.8) |
| In academy, institute or university | 114 | 44 (38.6) | 28 (14.6) | 18 (15.8) | 14 (12.3) | 10 (8.8) |
| In private lessons and/or tutoring | 191 | 103 (53.9) | 60 (31.4) | 17 (8.9) | 7 (3.7) | 4 (2.1) |
| Other: agriculture, labor, business, or other | 2051 | 954 (46.5) | 782 (38.2) | 203 (9.9) | 77 (3.9) | 35 (1.7) |
| **Recognition and socio-emotional support for teachers** | | | | | | |
| Psychological and/or emotional support (N = 10436) | | | | | | |
| If you received | 9476 | 4323 (45.6) | 3468 (36.6) | 1007 (10.6) | 429 (4.5) | 249 (2.6) |
| Did not receive | 9454 | 4171 (44.1) | 3540 (37.4) | 1048 (11.1) | 443 (4.7) | 252 (2.7) |
| Entity that provided psychological support and/or | | | | | | |
| DRE/UGEL/IE | 6760 | 3215 (47.6) | 2374 (35.1) | 706 (10.4) | 295 (4.4) | 170 (2.5) |
| Support from Family or friends | 2738 | 1203 (43.9) | 1055 (38.6) | 290 (10.6) | 128 (4.7) | 62 (2.3) |
| Private psychologist | 580 | 228 (39.3) | 227 (39.1) | 62 (10.7) | 35 (6.0) | 28 (4.8) |
| Ministry of health | 565 | 219 (38.8) | 256 (45.3) | 52 (9.2) | 21 (3.7) | 17 (3.0) |
| Other: EsSalud, others | 423 | 178 (42.1) | 165 (39) | 55 (13.0) | 17 (4.0) | 8 (1.9) |
| MINEDU - "I'm listening, teacher" | 301 | 143 (47.5) | 101 (33.5) | 37 (12.3) | 9 (3.0) | 11 (3.7) |
| Situations that a member of your household has experienced (N=10,436) | | | | | | |
| Lost or stopped working | 6350 | 2859 (45.0) | 2402 (37.9) | 570 (9.0) | 268 (4.2) | 151 (2.4) |
| Any illness other than COVID-19 | 6029 | 2461 (40.8) | 2425 (39.79) | 689 (11.4) | 279 (4.6) | 175 (2.9) |
| None | 4834 | 2171 (44.9) | 1715 (35.5) | 574 (11.9) | 233 (4.8) | 141 (2.9) |
| Diagnosed with COVID-19 | 4486 | 2031 (45.3) | 1736 (38.7) | 435 (9.7) | 183 (4.1) | 101 (2.3) |
| Have received any government benefits | 4155 | 2209 (53.2) | 1375 (33.1) | 357 (8.6) | 142 (3.4) | 72 (1.7) |
| Postponed or dropped out of school | 3673 | 1689 (46.0) | 1358 (36.9) | 367 (10.0) | 164 (4.5) | 95 (2.6) |
| Moved temporarily or permanently | 3177 | 1499 (47.2) | 973 (36.9) | 322 (10.1) | 118 (3.7) | 65 (2.0) |
| Did not respond | 80 | 21 (26.3) | 42 (48.8) | 12 (15.0) | 4 (5.0) | 4 (5.0) |
| Incentives that teachers should receive to promote innovations or good pedagogical practices (N=10,431) | | | | | | |
| Congratulatory resolutions | 1769 | 880 (49.7) | 574 (32.5) | 179 (10.1) | 82 (4.6) | 54 (3.1 |
| Monetary bonuses | 4440 | 1600 (36.0) | 1885 (42.4) | 567 (12.8) | 249 (5.6) | 139 (3.1) |
| Scholarships/Training | 12709 | 6007 (47.3) | 4545 (35.8) | 1308 (10.3) | 541 (4.3) | 308 (2.4) |
| They have developed or participated in a pedagogical innovation project or good pedagogical practice | | | | | | |
| Yes | 7453 | 3033 (40.7) | 2895 (38.9) | 894 (12.0) | 390 (5.2) | 241 (3.2) |
| No | 11459 | 5451 (47.6) | 4106 (35.8) | 1160 (10.1) | 482 (4.2) | 260 (2.3) |

*ICT= Information and Communication Technologies, IE = Educational Institution, DRE = Regional Directorate of Education, UGEL = Local Educational Management, MINEDU = Ministry of Education.

Table 3 shows the assessment of the teaching career using the instrument developed for this study. The categories were: high, medium, or low. Teachers described their career as highly valued for the following items: "occupying their time outside the workplace" (86.9%), "in-service training" (57.7%), and "receiving a stimulus (67.2%). They had medium assessments of: "type of occupation outside the workplace" (94.9%), "working conditions" (44.4%), and "psychological support" (50.1%). Low assessments were found for: "participation in a project" (60.6%) and "psychological support" (49.9%).

**Table 3.** Valuation of the teaching career, results of the 2020 National Teacher Survey

| Dimensions | High N (%) | Medium N (%) | Low N (%) |
|---|---|---|---|
| In-service training | 10927(57.7) | 5708(30.2) | 2291(12.1) |
| Working conditions | 10443(55.1) | 8408(44.4) | 99(0.5) |
| Occupying their time outside the workplace | 16455(86.9) | 2473(13.1) | N.A. |
| Type of occupation outside the workplace | N.A. | 2349(94.9) | 124(5.1) |
| Psychological support | N.A. | 9476(50.1) | 9454(49.9) |
| A family member's COVID-19 experience | 4486(23.7) | N.A. | 14444(76.3) |
| Receiving a stimulus conditions | 12709(67.2) | 4440(23.5) | 1769(9.3) |
| Participation in a project | 7453(39.4) | N.A. | 11459(60.6) |

*NA = Not applicable

Table 4 presents the results of the assessment of the future of the Public Teaching Career. The view with the highest percentage is that of conformism (48%), and the optimistic view is 2.2 percentage points lower.

**Table 4.** EBR teachers' vision of the future in the 2020 National Teacher Survey

| Types of vision | N (%) |
|---|---|
| Pessimistic vision of the future | 1178 (6.2) |
| Conformist vision of the future | 9097 (48.0) |
| Optimistic vision of the future | 8664 (45.8) |

Table 5 shows the bivariate analysis where categorized age, sex, professional career, vision of the future, and levels are related to the appreciation of the teaching career ($p<0.05$).

**Table 5.** Factors independently associated with the appreciation of the teaching career and multiple regression analysis with a generalized linear model

| Variables | Bivariate Analysis* | Multiple Regression* |
|---|---|---|

|  | PR | CI 95% | *p* | PR | CI 95% | *p* |
|---|---|---|---|---|---|---|
| Categorized age | | | | | | |
| 20 to 39 | Ref. | | | Ref. | | |
| 40 to 59 | 1,22 | 1.17 - 1.26 | <0.001 | 1,03 | 0.99 - 1.07 | 0,145 |
| 60 to 70 | 1,27 | 1.20 - 1.35 | <0.001 | 0,98 | 0.92 - 1.04 | 0,45 |
| Gender | | | | | | |
| Female | Ref. | | | Ref. | | |
| Male | 0,95 | 0.93 - 0.99 | 0,004 | 0,95 | 0.92 - 1.04 | 0,002 |
| Professional career | | | | | | |
| ale I | Ref. | | | Ref. | | |
| Scale II | 1,63 | 1.57 - 1.70 | <0.001 | 1,65 | 1.59 - 1.72 | <0.001 |
| Scale III | 1,63 | 1.56 - 1.70 | <0.001 | 1,64 | 1.57 - 1.72 | <0.001 |
| Scale IV | 1,62 | 1.55 - 1.70 | <0.001 | 1,64 | 1.56 - 1.72 | <0.001 |
| Scale V-VII | 1,67 | 1.57 - 1.78 | <0.001 | 1,68 | 1.57 - 1.79 | <0.001 |
| Hired | 1,71 | 1.59 - 1.85 | <0.001 | 1,72 | 1.59 - 1.86 | <0.001 |
| Vision of the future | | | | | | |
| Pessimistic | Ref. | | | Ref. | | |
| Conformist | 0,84 | 0.80 - 0.89 | <0.001 | 0,97 | 0.92 - 1.03 | 0,281 |
| Optimistic | 0,96 | 0.91 - 1.01 | 0,158 | 1,06 | 1.01 - 1.12 | 0,033 |
| Educational level | | | | | | |
| Preschool | Ref. | | | Ref. | | |
| Primary | 1,05 | 1.01 - 1.09 | 0,006 | 0,93 | 0.90 - 0.97 | <0,001 |
| Secondary | 1,03 | 0.99 - 1.07 | *0,153* | 1,01 | 0.97 - 1.05 | *0,576* |

*PR = Prevalence Ratio, CI = Confidence Interval

The instrument's reliability was assessed: internal consistency was identified with a Cronbach's alpha of 0.9313. Ideally, a value between 0.7 and 0.9 is indicated; a value exceeding 0.9 may indicate item redundancy or duplication. This suggests that a validation study is required to improve the instrument's validity and reliability; in this case, it would likely involve item reduction.

## 4. Discussion

This study used data on teachers in the Peruvian State, obtained through the Endo 2020 survey. Questions were used to analyze the perceptions of regular basic education teachers. Teacher characteristics regarding in-service training, working conditions and outlook, and socio-emotional recognition and support were identified. Teachers' values for the teaching career (pessimistic, conformist, and optimistic vision of the future) were identified, as well as factors associated with the appreciation of the teaching career. As a result of the multiple regression analysis, the following were associated with the appreciation of the teaching career: gender, professional career, a conformist vision of the future, and being a primary school teacher.

The Endo survey (2020) showed that the average age of Regular Basic Education (EBR) teachers is 45.4 years. The 40-59 age group represents the majority of teachers, more than half of whom are women. Most of these teachers work primarily in primary education, followed by those who teach in preschool and secondary education.

Gender appears to affect the evaluation of female teachers, with lower qualifications compared to males (Mengel et al., 2016). In another context, gender appears to influence satisfaction and organizational commitment among men and women (Apaza-Humerez &

Turpo-Chaparro, 2023). In the study, men were on average 45 years old, and their assessment of their future teaching career was considered conformist.

Regarding the teaching scale, these findings identify a high percentage of teachers without a teaching scale, which are those teachers who do not have a degree, are high school graduates, graduates, or university students. Another high percentage is found in groups I and II of the teaching scale level. The 2001 study by Díaz and Saavedra mentions that the teaching scale is key in career evaluation and helps standardize salaries (Díaz and Saavedra, 2001). Thus, better compensation linked to the teaching scale improves satisfaction and even teacher retention rates (Huaita et al., 2024).

In this study, when measuring the "In-service training" dimension, we noted that it is higher in the group without a scale and in levels I and II compared to the highest levels of the teaching scale (Pérez et al., 2023). Teachers at the highest levels rarely participate in training provided by the Ministry of Education because they have other academic priorities (Pérez et al., 2023), a situation also observed in the results. But we can mainly affirm that among teachers without a scale and in levels I-II combined, less than 50% participate in virtual courses, conferences, or seminars, and ICT training. This point can be leveraged to continue offering innovative training aimed at improving teachers' capacity and thus increasing the participation rate.

Another factor is ongoing training and professional development, which are relevant aspects of revaluing the teaching profession. Training in new educational methodologies and technologies improves teachers' competencies (Castillo, 2023; Taimal and Gonzáles, 2022). Providing training and professional development strengthens teachers by making them confident, enhancing their skills, and increasing their confidence in their roles (Vélez Belisario et al., 2024).

The revaluation of the teaching profession can be based on theories or approaches, such as the meritocratic model, which seeks to improve educational quality and professional recognition for teachers. The aforementioned theoretical framework has already been implemented in several educational contexts and in countries such as Mexico. The model establishes the promotion and recognition of teaching staff based on standardized evaluations with a multifactorial assessment, including qualitative and quantitative indicators (Cordero-Arroyo et al., 2021; Paz and Gómez, 2021). This structure seeks to ensure teacher qualifications and foster a healthy, competitive environment to improve educational practice.

The European Qualifications Framework offers clear standards for teacher training and assessment (Echeverria et al., 2020; Echeverria & Farran, 2018), thus aligning with the proposal for competency-based training. Teachers' adoption of competencies is key in education to generate a methodological change that promotes knowledge transfer, active learning, and student engagement (Bustamante & Elera, 2023; Castillo, 2023).

Another relevant factor in the revaluation of the teaching career for authors such as Sierra (2019) and Yoza and Velez (2021) is intrinsic motivation through the application of active methodologies. Furthermore, educational policies promote competency-based training. This has been decisive in achieving a revaluation of the teaching profession, which must continue to improve and achieve higher optimistic appreciation.

Other factors related to satisfaction or dissatisfaction include economic conditions, professional recognition, and the work environment. Despite new laws and investment, many teachers in Peru express dissatisfaction due to unrecognized contributions and difficult working conditions (Maldonado-Cueva, 2024; Holgado Apaza et al., 2023).

In short, training, capacity building, and institutional support allow for an increase in the value of the profession, and regarding remuneration, income is an indicator of job satisfaction (Díaz & Savedra, 2001; Huaita et al., 2024; Holgado Apaza et al., 2023).

Finally, although age has not been associated with the valuation of the teaching career, according to several authors, being young and a teacher favors its qualification (Lobos and Cortés, 2023). Being older can be favorable due to the experience gained and the ability to make decisions; conversely, it can be a limitation, both physical and cognitive (Barrantes, 2021; Yllesca et al., 2024). In this sense, age may be related to how the teaching career is valued; in our study, being male and having an average age of 45 years is a factor that is related to the valuation of the teaching career.

## 5. Conclusions

In conclusion, EBR teachers primarily have a conformist and optimistic view of the future. The factors associated with the appreciation of the teaching career, measured through a GLM model, were being male, belonging to ranks I, II, III, IV (V-VII), and teaching in primary education (p<0.005). The appreciation of the teaching career is based on a comprehensive approach that combines meritocracy, competency-based training, intrinsic motivation, and continuous professional development. These elements are crucial to ensuring that teachers are not only recognized for their work but also prepared to face the challenges of an ever-evolving educational environment. Reflecting on teaching practice aims to continue encouraging learning in those already at a certain level in their professional careers, thereby improving and increasing their level. The effect of mentioning optimistic appreciation strengthens teachers' professional identity and leads to an improvement in education.

**Author Contributions:** (CRedIT)

"Conceptualization, M.L-U. and D.H.A; methodology, M.L-U. and R.B.A; software, .F.L.C.; formal analysis, F.L.C.; investigation, M.L-U; M.B.A.; resources, X.X.; data curation, F.L.C.; writing—original draft preparation, M.L-U. and D.H.A; writing—review and editing, R.F-A.; visualization, R.F-A. and R.B.A; supervision, M.L-U.; R.F-A. project administration, M.L-U. All authors have read and agreed to the published version of the manuscript."

**Funding:** This research received no external funding

**Abbreviations**

The following abbreviations are used in this manuscript:

| | |
|---|---|
| PEN | National Education Project |
| ENDO | National Survey of Teachers |
| GLM | Generalized Linear Model |
| MINEDU | Ministry of Education |
| PISA | International Student Assessment |
| EBR | Regular Basic Education |
| DRE | Regional Directorate of Education |
| UGEL | Local Educational Management |
| IE | Educational institution |
| ICT | Information and Communication Technologies |